\documentclass[aip,pof,preprint]{revtex4-1}
\usepackage{amssymb,latexsym, amsmath,graphicx,bm}
\usepackage{geometry}

\newcommand{\wh}[1]{{\widehat{#1}}}
\newcommand{\bfsigma}{\mathbf \sigma}
\newcommand{\be}{\begin{equation}}
\newcommand{\ee}{\end{equation}}
\newcommand{\vv}[1]{{\mathbf #1}}
\newcommand{\eqn}[1]{\begin{equation}#1\end{equation}}
\newcommand{\ls}[1]{#1_l}
\newcommand{\remove}[1]{}

\begin{document}
\bibliographystyle{unsrtnat}

\title{
The Equivalence of the Lagrangian-Averaged Navier-Stokes-$\alpha$ Model and the Rational LES model in Two Dimensions
}
\author{Balasubramanya T. Nadiga}
\email{balu@lanl.gov}
\affiliation{LANL, Los Alamos, NM-87545}
\author{Freddy Bouchet}
\email{freddy.bouchet@ens-lyon.fr}
\affiliation{ENS-Lyon, CNRS, Lyon, France}
\date{\today}
\begin{abstract}
  In the Large Eddy Simulation (LES) framework for modeling a
  turbulent flow, when the large scale velocity field is defined by
  low-pass filtering the full velocity field, a Taylor series
  expansion of the full velocity field in terms of the large scale
  velocity field leads (at the leading order) to the nonlinear
  gradient model for the subfilter stresses. Motivated by the fact
  that while the nonlinear gradient model shows excellent a priori
  agreement in resolved simulations, the use of this model by itself
  is problematic, we consider two models that are related, but better
  behaved: The Rational LES model that uses a sub-diagonal Pade
  approximation instead of a Taylor series expansion and the
  Lagrangian Averaged Navier-Stokes-$\alpha$ model that uses a
  regularization approach to modeling turbulence. In this
  article, we show that these two latter models are identical in two
  dimensions.
\end{abstract}


\maketitle

\section{Introduction}
In a turbulent flow, it is usually the case that energy is
predominantly contained at large scales where as a disproportionately
large fraction of the computational effort is expended on representing
the small scales in fully-resolved simulations of such flows (e.g.,
see Pope, 2000~\cite{pope-00}).  Large eddy simulation (LES) is a
technique that aims to explicitly capture the large, energy-containing
scales while modeling the effects of the small scales that are more
likely to be universal. This technique is both popular and is by far
the most successful approach to modeling turbulent flows. We note,
however, that in complex, wall-bounded and realistic configurations
(such as, e.g., encountered in industrial situations), computational
requirements for LES is still prohibitive that a hybrid (Reynolds
Averaged Navier Stokes) RANS-LES approach is favored.\cite{rajamani-10}

The nature of the dynamics of large scale circulation in the world
oceans and planetary atmospheres is quasi two dimensional due to
constraints of geometry (small vertical to horizontal aspect ratio),
rotation and stable stratification. For example, consider the
(inviscid and unforced) quasi-geostrophic equations that describe the
dynamics of the large, geostrophically and hydrostatically balanced,
scales:
\be
{D q \over D t} = {\partial q \over \partial t} + \vv u\cdot \nabla q = 0 \ee
where $q$ is potential vorticity approximated in the quasi-geostrophic limit by
\be
q=\nabla^2\psi + {\partial \over \partial z}\left({f_0^2\over
    N^2}{\partial \psi \over \partial z}\right) +\beta y,
\label{qgpv}
\ee
and $\vv u$ is the advection velocity approximated in the quasi-geostrophic limit by the
geostrophic velocity given in terms of streamfunction
$\psi$ by $\vv u \approx \vv u_g=\vv k \times \nabla \psi$.
In other notation, $\nabla$ is the horizontal gradient
operator, $f$ is the Coriolis parameter given by $f = f_0 + \beta y$
in the $\beta$-plane approximation, $y$ is the meridional coordinate, $N^2$ is the Brunt-Vaisala
frequency given in terms of the specified density gradient by $N^2=
{g\over \rho_0}{d\rho \over dz}.$ On the one hand, the particle-wise advection of
potential-vorticity, and the dual conservation of (quadratic
quantities) energy and potential-enstrophy are properties shared by
quasi-geostrophic dynamics in common with two-dimensional flows.
On the other hand, quasi-geostrophic dynamics shares in common with three-dimensional flow,
the property of vortex stretching (in this limit, it is only the planetary vorticity $f_0$ that is
stretched and is represented by the $\partial/\partial z$ term in (\ref{qgpv}).

It is the qualitative similarity of turbulence in these systems with
two-dimensional turbulence, as elucidated by Charney, 1971,~\cite{charney-71} that is the
primary reason for interest in two-dimensional turbulence. The dual
conservation of (potential) enstrophy and energy in (quasi)
two-dimensional turbulence leads to profound differences as compared
to fully three-dimensional turbulence: there exist two inertial
regimes---a forward-cascade of (potential) enstrophy regime and an
inverse-cascade of energy regime---in (quasi) two-dimensional
turbulence in contrast to the single forward-cascade of energy regime
in fully three-dimensional turbulence.

In the context of LES, which aims to model the effects of
small-scales, it is clearly the forward-cascade inertial regimes that
are of direct relevance.  One of the most popular LES model is the
Smagorinsky model,\cite{smag-63} and this class of eddy-viscosity models assume that
the main effect of the unresolved scales is to remove, from the
resolved scales, either energy for 3D flows or (potential) enstrophy
for (quasi-geostrophic) 2D flows---the appropriate quantity that is
cascading forward. However, an examination of the statistical
distribution of the transfer of either energy in 3D turbulence or
(potential) enstrophy in (quasi-geostrophic) 2D turbulence in the
forward cascade regime\citep[see][]{chen-03,nadiga-08, nadiga-09}
demonstrates that the net forward cascade results from the
forward-scatter being only slightly greater than the 
backscatter. Clearly, models such as the Smagorinsky model, or more
generally scalar eddy-viscosity, by modeling only the net
forward cascade, fail to represent possible important dynamical
consequences of backscatter.

The recent reinterpretation of the classic work of Leray---which
considered a mathematical regularization of the advective
nonlinearity---in terms of the LES formalism,
has given rise to the so-called regularization approach to modeling
turbulence.$^{\hbox{e.g., }}$ \cite{geurts-06} An important model in this approach is the
Lagrangian-Averaged Navier-Stokes-$\alpha$ (LANS-$\alpha$) model
introduced by Holm and co-workers.\cite{holm-98}

The origins of the LANS-$\alpha$ turbulence model lie in 1) the notion
of averaging over a fast turbulent spatial scale $\alpha$, the
reduced-Lagrangian that occurs in the Euler-Poincare formalism of
ideal fluid dynamics~\cite{holm-98}, and in 2) three-dimensional
generalizations~\cite{holm-98} of a nonhydrostatic shallow water
equation system, known in literature as the Camassa-Holm
equations.~\cite{chen-98} However, viewed from the point of view of
the regularization approach, this model can be thought of as a
particular frame-indifferent (coordinate invariant) regularization of
the Leray type that preserves other important properties of the
Navier-Stokes equations such as having a Kelvin theorem. To add to the
richness of this model, almost exactly the same equations arise in the
description of second-grade fluids~\cite{noll-65,graham-08} and vortex-blob
methods.$^{\hbox{e.g., }}$ \cite{nadiga-01a}

There is now an extensive body of literature covering various aspects
of the LANS-$\alpha$ model. In particular with respect to its
turbulence modeling characteristics, analytical computation of the
model shear stress profiles has shown favorable comparisons against
laboratory data of turbulent pipe and channel flows~\cite{chen-98} and
a posteriori comparisons of mixing in three-dimensional temporal
mixing layer settings,~\cite{geurts-06} in isotropic homogeneous
turbulence settings,~\cite{chen-99,mohseni-03} and in anisotropic
settings~\cite{zhao-05, scott-09} compare well against Direct
Numerical Simulations (DNS). In three dimensions, it has, however,
recently been noted~\cite{graham-08} that the use of LANS-$\alpha$
model as a subgrid model can be deficient in certain respects.  In the
two dimensional and quasi-two dimensional contexts, a posteriori
comparisons of LANS-$\alpha$ based computations have shown favorable
comparisons against eddy-resolving
computations.~\cite{nadiga-01b,holm-03,hecht-08} Nevertheless, this
model has mostly been viewed as a complementary approach to modeling
turbulence.

The nonlinear gradient model~\cite{leonard74,clark79,meneveau-00} and the Rational LES
model~\cite{galdi-00,iliescu03,berselli05} are part of another class of LES models, built on
a direct dynamical analysis of what should be a good approximation of
the effect of the subgrid scales on the largest scales, through
turbulent stresses. When the large scale velocity field is defined by
low-pass filtering the velocity field, a natural asymptotic expansion
leads to approximated turbulent stresses. This defines the nonlinear
gradient model. An essential point is that the actual turbulent
stresses of 2D and quasi-geostrophic turbulent flows, computed from
direct numerical simulation, have been shown to be well approximated
by the one defining the nonlinear gradient
model.~\cite{bouchet01,chen-03, nadiga-08,nadiga-09}

The nonlinear gradient model (\ref{nl-model}) uses a natural
approximation of the turbulent stresses. However this model has
several drawbacks. Indeed, whereas it has been proven that the
nonlinear gradient model turbulent stress (\ref{nl-model}) preserves
energy for two dimensional flows,~\cite{bouchet01} this is
generally not the case in three dimensional flows, and instabilities
or finite time energy blow up can occur. The situation is not much
better in 2D and quasi-geostrophic flows in that the incompressiblity
constraint implies that the divergence of the deformation tensor
($\sigma$ in equation (\ref{nl-model})) generally has a positive
definite direction and a negative definite direction.  Physically,
this amount to an anistropic viscosity with positive value in some
directions and negative values in other
directions.~\cite{bouchet01, chen-03,
  nadiga-08,nadiga-09} These drawbacks mean that the nonlinear model
is not a good physical model and will lead to instabilities, for two
dimensional, quasi-geostrophic and three dimensional flows. An
alternative model based on entropic closures, keeping the main
properties of the nonlinear model (good approximation of the turbulent
stresses, conservation of energy), has been proposed and proven to
give very good results for two-dimensional flows.
\cite{bouchet01} In three dimensions, Domaradzki and Holm,
2001,~\cite{domaradzki-01} note that one component of the
LANS-$\alpha$ (subfilter stress) model corresponds to the subfilter
stress that would be obtained upon using an approximate deconvolution
procedure on the nonlinear gradient model.

Analysis of the drawbacks of the nonlinear gradient model led Galdi
and Layton to propose the Rational LES model.~\cite{galdi-00} The Rational LES
model coincides with the nonlinear model at leading order, but
provides a stronger attenuation of the smallest scale. As confirmed by
recent mathematical results,~\cite{berselli05} the Rational LES model is
well posed and should lead to stable numerical algorithms. It is thus
a good candidate for LES. 

The Nonlinear-Gradient model has been well studied over more than
three decades. These studies started with Leonard, 1974~\cite{leonard74} and Clark,
Ferziger and Reynolds, 1979.~\cite{clark79}  Rather than attempt an incomplete survey
of the literature relevant to the a priori and a posteriori testing of
this model here, we note that a fairly modern account of this can be
found in Meneveau and Katz, 2000.\cite{meneveau-00} The more recent aspect of the
Rational LES model is in making the highly favorable a priori
comparisons of the Nonlinear-Gradient model more amenable to a
posteriori simulations. For example, Iliescu et al., 2003\cite{iliescu03}
compare the behavior of the Rational LES model to the Nonlinear
Gradient model (and the Smagorinski model) in the 2D and 3D cavity
flow settings, both at low and high Reynolds numbers. They find a)
that laminar flows are correctly simulated by both models, and b) that
at high Reynolds numbers, the Nonlinear Gradient model simulations,
either with or without the Smagorinski model, lead to a finite time
blow-up while the Rational LES model simulation displayed no such
problem and succeeded in its LES role, i.e., compared to a fine-scale
resolved simulation, the Rational LES model was able to capture and
model the large-eddies well on a coarse mesh. Furthermore, they find
that the Rational LES model performed better than the Smagorinski
model alone in capturing the behavior of the large-eddies.  Finally,
we note that with both the Rational LES model and the LANS-$\alpha$
models, the burden of modeling borne by the additional dissipative
term is smaller than in other approaches.

Following the development of these models, we note that the
LANS-$\alpha$ and the rational LES model have interesting
complementary properties: While the LANS-$\alpha$ preserves the Euler
equation structure through the Kelvin theorem, the Rational LES model
develops a good approximation of the turbulent stresses while
ameliorating problems associated with the nonlinear gradient model. It
would thus be useful to examine the relation between these two models.
In this article, we demonstrate the equivalence of the LANS-$\alpha$
to the Rational LES model in two-dimensions.  By equivalence, we mean
here that, the evolution equations for one of the models can be
exactly transformed into the other. As will be evident, given the very
different approaches taken in arriving at these models, it'll involve
more than a simple transformation; it will also involve disentangling
the turbulence term implied by the particular regularization of the
nonlinear term. The importance of this result lies in the fact that
mathematical results obtained for one of these models become also true
for the other. We also demonstrate that these two models are different
in three dimensions.

In sections 1 and 2, after recalling the framework of turbulent
stresses and LES, we briefly describe the nonlinear gradient, the
Rational LES, and the LANS-$\alpha$ models. In section 3, we prove the
equivalence of the Rational LES and of the LANS-$\alpha$ models in two
dimensions. In section 4, we prove that they are not equivalent in
three dimensions. After a brief numerical example, implications of the above results are discussed in
the final  section.

\section{LES of Two-Dimensional Turbulence and the nonlinear-gradient model}
In LES, the resolution of energy containing
eddies that dominate flow dynamics is made computationally feasible by
introducing a formal scale separation.~\cite{pope-00} The scale separation is
achieved by applying a low-pass filter \textbf{$G$} with a characteristic scale $\alpha$ ($2\alpha^2$ is the second moment of $G$) to the original equations.
To this end, let the
fields $\vv u$, $q$, {\it etc...} be split into large-scale (subscript $l$)and small-scale (subscript $s$)
components as
$$ \vv u 
= \vv u_l + \vv u_s,$$
where
$$ 
\vv u_l (\vv x) = \int_D G(\vv x - \vv x') \vv u (\vv x') d\vv x' ,
$$
$$
\vv u_s (\vv x) = \vv u -\vv u_l,$$
the filter function $G$ is normalized so that
$$ \int_D G(\vv x') d\vv x' =1,$$
and where the integrations are over the full domain $D$. In contrast to
Reynolds decomposition, however, generally, $\vv u_{ll} \neq \vv u_l$ and
$\vv u_{sl} \neq 0$.

For convenience, we write the two-dimensional vorticity equation as
\eqn{ {D\omega\over Dt}={\partial \omega \over \partial t} + \vv
  u\cdot \nabla \omega = \nabla\times F_{2d}+\nabla\times D_{2d} = F +
  D,
\label{2d-vort}}
where $F_{2d}$ is the two-dimensional momentum forcing, $D_{2d}$ is dissipation, and where, for brevity, we denote
$\nabla\times F_{2d}$ by $F$, and $\nabla\times D_{2d}$ by $D$. Applying the filter to (\ref{2d-vort})
 leads to an equation for the
evolution of the large-scale component of vorticity which is the primary
object of interest in LES:
\eqn{
{\partial \ls{\omega}\over\partial t}
+ \nabla\cdot\left(\ls{\vv u}\ls{\omega} \right)
= F_l + D_l
- \nabla\cdot\bfsigma
\label{eqnles}
}
where
\eqn{
\bfsigma=\ls{\left(\vv u \omega\right)} - \ls{\vv u}\ls{\omega}
\label{totalsgs}
}
is the turbulent sub-filter vorticity-flux, and as in (\ref{2d-vort}), we denote
$\left(\nabla\times F_{2d}\right)_l$ by $F_l$, and so also for dissipation. This turbulent subgrid
vorticity-flux may in turn be written in terms of the Leonard stress,
cross-stress, and Reynolds stress~\cite{pope-00} as
\eqn{
\sigma=
 \underbrace{(\vv u_l \omega_l)_l - \vv u_l \omega_l}_{\hbox{Leonard stress}}
+\underbrace{(\vv u_l \omega_s)_l + (\vv u_s \omega_l)_l}_{\hbox{Cross-stress}}
\quad\quad+\underbrace{(\vv u_s \omega_s)_l}_{\hbox{Reynolds stress}}.
\label{gvarsgs}
}
However, while $\sigma$ itself is Galilean-invariant,
the above Leonard- and Cross-stresses are not Galilean-invariant.
Thus when these component stresses are considered individually, the following
decomposition, originally due to Germano, 1986\cite{germano-86} is preferable
\eqn{
\sigma=
 \underbrace{(\vv u_l \omega_l)_l - \vv u_{ll} \omega_{ll}}_{\hbox{Leonard Stress}}
+\underbrace{(\vv u_l \omega_s)_l + (\vv u_s \omega_l)_l - \vv u_{ll} \omega_{sl} -\vv u_{sl} \omega_{ll}}_{\hbox{Cross-stress}}
+\underbrace{(\vv u_s \omega_s)_l-\vv u_{sl} \omega_{sl}}_{\hbox{Reynolds stress}}.
\label{ginvarsgs}
}

The filtered equations, which are the object of simulation on a grid
with a resolution commensurate with the filter scale in LES, are then closed
by modeling subgrid-scale (SGS) stresses to account for the effect of
the unresolved small-scale eddies. In this case (\ref{eqnles}) will be
closed on modeling the turbulent subgrid vorticity-flux $\bfsigma$.

As is tradition, a Gaussian filter is chosen.
In eddy-permitting simulations, some of the range of scales of
turbulence are explicitly resolved. Therefore, information about the
structure of turbulence at these scales is readily available. In
LES formalism, there is a class of models that attempt to model the smaller
unresolved scales of turbulence based on
the assumption that the structure of the turbulent velocity
field at scales below the filter scale is the same as the structure of
the turbulent velocity field at scales just above the filter scale.
\cite{meneveau-00}


Further expansion of
the velocity field in a Taylor series and performing filtering
analytically results in
\eqn{
\ls{(u_i u_j)} \propto {\partial u_{li} \over \partial x_k}
{\partial u_{lj} \over \partial x_k},
\label{nlmdl}
}
a quadratic nonlinear combination of resolved gradients
for the subgrid model.~\cite{leonard74,clark79}
The
interested reader is referred to Meneveau and Katz, 2000\cite{meneveau-00} for a comprehensive
review of the nonlinear-gradient model.

Equivalently, expansion of $\vv u_l$ and $\omega_l$ in the Galilean-invariant form
of the Leonard-stress component of the sub-filter
eddy-flux of vorticity (\ref{ginvarsgs}) in a Taylor series:
$$(\vv u_l \omega_l)_l - \vv u_{ll} \omega_{ll}=$$
$$=\int d\vv x' \;G(\vv x -\vv x')
\left(\vv u_l(\vv x) + (x' - x)_j
{\partial u_{li} \over \partial x_j}(\vv x)\right)
\left(\omega_l(\vv x) + (x' - x)_j
{\partial \omega_{l} \over \partial x_j}(\vv x)\right)-$$
$$\int d\vv x' \;G(\vv x -\vv x') \left(\vv u_l(\vv x) + (x' - x)_j
{\partial u_{li} \over \partial x_j}(\vv x)\right)*
\int d\vv x' \;G(\vv x -\vv x') \left(\omega_l(\vv x) + (x' - x)_j
{\partial \omega_{l} \over \partial x_j}(\vv x)\right)
$$
%
produces at the first order
\eqn{
\sigma
=2\alpha^2{\partial u_{li} \over
  \partial x_j}\; {\partial \omega_l \over \partial x_i} + \mathcal{O}(\alpha^4)
=2 \alpha^2 \nabla\vv u_l \cdot \nabla \omega_l + \mathcal{O}(\alpha^4),
\label{sigma-nl-model}
}
where $2\alpha^2$ is the second moment of the filter used. 
The leading order is again a quadratic nonlinear combination of resolved gradients. The approximate model that retains only the second order term is called the nonlinear gradient model. In this two-dimensional setup, it reads
\begin{equation}
\frac{\partial\omega_l}{\partial t}+\mathbf{u}_l\boldsymbol{\cdot\nabla}\omega_l=-2\alpha^{2}\left[\nabla\mathbf{u}_l^{T}.\nabla(\nabla\omega_l)\right] + F_l + D_l
\label{eq:nl-model-2D}
\end{equation}
(please see the appendix for the definition of operator $\nabla\mathbf{u}_l^{T}.\nabla$).
 
 For simplicity, we have presented the two-dimensional derivation of the nonlinear gradient model, however similar considerations lead to the three dimensional nonlinear gradient model:
\begin{equation}
  \frac{\partial\vv{u}_l}{\partial t}+\vv{u}_l\boldsymbol{\cdot\nabla}\vv{u}_l =-2\alpha^{2}\boldsymbol{\nabla\cdot}\left[\nabla\vv{u}_l\nabla\vv{u}_l\right]-\boldsymbol{\nabla}P + \left(F_{3d}\right)_l + \left(D_{3d}\right)_l.
\label{nl-model}
\end{equation}
In the two dimensional context, this model has been derived by Eyink, 2001
\cite{eyink-01} without the self-similarity assumption, but rather by
assuming scale-locality of contributions to $\bfsigma$ at scales
smaller than the filter scale, and its use has been investigated by
various authors.~\cite{bouchet01,chen-03} Nadiga, 2008 and
2009\cite{nadiga-08,nadiga-09} have demonstrated excellent {\em a
  priori} testing of the nonlinear gradient model in quasi-geostrophic
turbulence, the same also holds in the three-dimensional turbulence
context.$^{\hbox{e.g., }}$ \cite{meneveau-00} The nonlinear gradient
model, however, holds much better in two-dimensional and quasi
two-dimensional settings than in fully three-dimensional settings.

\section{Rational LES model and the LANS-$\alpha$ model}

\subsection{Rational LES model}
By analyzing the nonlinear-gradient model in terms of Fourier
components, Galdi and Layton, noted that the nonlinear-gradient model
increases the high wavenumber components (scales smaller than the
filter scale) and therefore does not ensure that $\omega_l$ is
smoother than $\omega$. Consequently, to remedy this problem, they
proposed an approximation which attenuates the small scale
  eddies, but is of the same order accuracy for large eddies (the two
  approximations coincide at order $\alpha^2$, see (\ref{nl-model})).

To this end, rather than using a Taylor expansion of the filter
($e^{-b x^2} \approx 1 - b x^2$), they considered the rational
approximation
\be
e^{-b x^2} \approx {1\over 1+b x^2}
\ee
Using the above sub-diagonal Pade approximation, the modified
nonlinear-gradient model leads to the 'Rational LES' model. We refer
to Galdi and Layton, 2000\cite{galdi-00} for the derivation of the evolution equation for
$\vv{u}_l$ (which is an approximation of the large scale component of
the full velocity field $\vv u$.)  It is
\begin{equation}
  \frac{\partial\vv{u}_l}{\partial t}+\vv{u}_l\boldsymbol{\cdot\nabla}\vv{u}_l =-2\alpha^{2}\left(I-\alpha^{2}\Delta\right)^{-1}\boldsymbol{\nabla\cdot}\left[\nabla\vv{u}_l\nabla\vv{u}_l\right]-\boldsymbol{\nabla}P + \left(F_{3d}\right)_l + \left(D_{3d}\right)_l,\label{eq:rational-velocity}\end{equation}
with $\boldsymbol{\nabla \cdot} \vv{u}_l = 0$, and where $\left(I-\alpha^{2}\Delta\right)^{-1}$ is the inverse of the operator $\left(I-\alpha^{2}\Delta\right)$ (easily expressed in a Fourier basis).

\subsection{The LANS-$\alpha$ model}

In the context of the three-dimensional incompressible Navier-Stokes
equations
\eqn{
{\partial \vv u \over \partial t}
+ {\vv u}\cdot\nabla\vv  u
= -\nabla \phi + F_{3d}+\nu\Delta \vv u ~~~;~~~ \nabla\cdot\vv u = 0,
\label{ins}
}
on a suitable domain with
appropriate boundary conditions, Leray regularization of (\ref{ins}) is  expressed by $^{\hbox{e.g., }}$ 
\cite{geurts-06}
\eqn{ {\partial \vv u \over \partial t} + {\vv u_l}\cdot\nabla\vv u =
  -\nabla \phi + F_{3d}+\nu\Delta u,
\label{leray}
}
where ${\vv u_l}$ is the large scale component of velocity filtered at
a characteristic length $\alpha$, $\phi=p/\rho_0$ is the normalized
pressure, $F_{3d}$ is the external forcing and $\nu$ the kinematic
viscosity. The filtered velocity ${\vv u_l}$ can be obtained by
application of a convolution filter to $\vv u$. A particularly
important example is the Helmholtz filter, to which we turn
momentarily. The Leray approach is basic to many recent studies in
regularized turbulence.  This regularization model does not preserve
some of the properties of the original equations (\ref{ins}), such as
a Kelvin circulation theorem. This is where the LANS-$\alpha$
formulation provides an important extension.

  A transparent way to present the LANS-$\alpha$ model is obtained
  when the incompressible Navier-Stokes (momentum) equations are
  written in the equivalent form
\eqn{
{\partial \vv u \over \partial t}
- {\vv u}\times\left(\nabla \times \vv  u \right)
= -\nabla \phi + F_{3d}+\nu\Delta \vv u.
\label{inscurl}
}
The LANS-$\alpha$ model is then given by $^{\hbox{e.g., }}$ \cite{holm-98,geurts-06}
\eqn{
{\partial \vv u \over \partial t}
- {\vv u_l}\times\left(\nabla \times \vv  u \right)
= -\nabla \phi + F_{3d}+\nu\Delta \vv u.
\label{insacurl}
}
Thus, just as the Leray regularization corresponds to the filtering
of the advecting velocity, the LANS-$\alpha$ regularization amounts to
filtering the velocity in the nonlinear term when written as the
direct product of a velocity and a vorticity {${\bm{\omega} = \nabla
    \times \vv u}$}. The LANS-$\alpha$ model may be written in the more common advective nonlinearity form
\eqn{
\frac{\partial \vv{u}}{\partial t} +
\vv{u_l}\cdot\nabla\vv{u}
- \alpha^2 \left(\nabla{\vv u_l}\right)^T\Delta {\vv u_l}
=- \nabla p
+F_{3d}+\nu\Delta\vv{u}.
\label{insa}
}
The filtered velocity is obtained by an inversion of the
Helmholtz operator: ${\vv
  u_l}{=(1-\alpha^2\Delta)^{-1}\vv u}$ with appropriate boundary
conditions. (It has to be noted that in a non-periodic domain, the boundary
  conditions that are necessary to invert the Helmholtz operator are
  specific to this modeling procedure.) The third term on the left in (\ref{insa}) is
introduced in the LANS-$\alpha$ modeling approach to restore a Kelvin
theorem to the modeled equations.

It is also instructive to consider the evolution of vorticity. For the Navier-Stokes equation, vorticity evolution is
\eqn{
\frac{\partial \bm{\omega}}{\partial t} +
\vv{u}\cdot\nabla\bm{\omega} =
\bm{\omega}\cdot\nabla\vv u + \nabla\times F_{3d} +\nu\Delta\bm{\omega}.
\label{ns-vort}
}
The vorticity evolution corresponding to the LANS-$\alpha$ model is
\eqn{
\frac{\partial \bm{\omega}}{\partial t} +
\vv{u_l}\cdot\nabla\bm{\omega} =
\bm{\omega}\cdot\nabla\vv u_l + \nabla\times F_{3d} +\nu\Delta\bm{\omega},
\label{ns-vort-a}
}
where in addition to a filtered advecting velocity, a mollification of
the vortex-stretching term is evident.

In two dimensions, (\ref{ns-vort-a}) reduces to
\eqn{
\frac{\partial {\omega}}{\partial t} +
\vv{u_l}\cdot\nabla{\omega} =
\nabla\times F_{2d} + \nabla\times D_{2d} = F + D,
\label{vort2d-a}
}
where forcing and dissipation have been written in correspondance with
the notation used in the two dimensional vorticity equation
(\ref{2d-vort}) and its LES counterpart (\ref{eqnles}).

\section{Identity of the Rational LES and LANS-$\alpha$ models in two dimensions}

In this section, we consider the Rational LES model and the
LANS-$\alpha$ models in two-dimensions.  Taking the curl of the
two-dimensional velocity equation for the Rational LES model
(\ref{eq:rational-velocity}), we obtain the vorticity equation
\begin{equation}
  \frac{\partial\omega_l}{\partial t}+\mathbf{u}_l\boldsymbol{\cdot\nabla}\omega_l=-2\alpha^{2}\left(I-\alpha^{2}\Delta\right)^{-1}\left[\nabla\mathbf{u}_l^{T}.\nabla(\nabla\omega_l)\right] + F_l + D_l,\label{eq:rational}\end{equation}
where $\omega_l$ is the vertical component of $\bm\omega_l$.
In order to compare the Rational LES model (\ref{eq:rational}) with the LANS-$\alpha$ model (\ref{vort2d-a}), we apply operator $\left(I-\alpha^{2}\Delta\right)$ to (\ref{eq:rational}) and
write the evolution equation for $\omega$ as 
\begin{equation}
\frac{\partial\omega}{\partial t}+\mathbf{u}_l\boldsymbol{\cdot\nabla}\omega= \delta M + F + D.
\label{eq:evolution-M}
\end{equation}
Comparing (\ref{eq:evolution-M}) with (\ref{vort2d-a}), we note that
$\delta M$ is the difference between the two models and is given by
\[
\delta M=-2\alpha^{2}\left[\nabla\mathbf{u}_l^{T}.\nabla(\nabla\omega_l)\right]+\mathbf{u}_l\boldsymbol{\cdot\nabla}\left[\left(I-\alpha^{2}\Delta\right)\omega_l\right]-\left(I-\alpha^{2}\Delta\right)\left[\mathbf{u}_l\boldsymbol{\cdot\nabla}\omega_l\right].\]
By direct computation
this expression simplifies to 
\[ \delta M=\alpha^{2}\left\{
  -2\left[\nabla\mathbf{u}_l^{T}.\nabla(\nabla\omega_l)\right]-\mathbf{u}\boldsymbol{\cdot\nabla}\left[\Delta\omega_l\right]+\Delta\left[\mathbf{u}_l\boldsymbol{\cdot\nabla}\omega_l\right]\right\}
.\] 
Then using the vector calculus identity (\ref{eq:Identity3}) in the appendix, we conclude that $\delta M=0$. The
dynamics of $\omega$ is thus the same as given by the $LANS-\alpha$ model\[
\frac{\partial\omega}{\partial
  t}+\mathbf{u}_l\boldsymbol{\cdot\nabla}\omega=F + D.\] We thus conclude
that the Rational LES model and the LANL-$\alpha$ models are equivalent in two
dimensions.

\section{The Rational LES and LANS-$\alpha$ models are different in three dimensions}

In three dimensions, the Rational LES model for an incompressible flow ($\nabla.\mathbf{u}=0$) is \begin{equation}
\frac{\partial\mathbf{u}_l}{\partial t}-\mathbf{u}_l\boldsymbol{\times}\left(\nabla\boldsymbol{\times}\mathbf{u}_l\right)=-\frac{\mbox{\ensuremath{\nabla}}P_{1}}{\rho}-2\alpha^{2}\left(I-\alpha^{2}\Delta\right)^{-1}\left[\nabla\mathbf{u}_l^{T}.\nabla(\nabla\mathbf{u}_l)\right] + \left(F_{3d}\right)_l + \nu\Delta\vv u_l,\label{eq:rational_model_3D}\end{equation}
where $P_{1}$ is the sum of the physical and kinetic pressure. The
LANS-$\alpha$ model is \begin{equation}
\frac{\partial\mathbf{u}}{\partial t}-\mathbf{u}_l\boldsymbol{\times}\left(\nabla\boldsymbol{\times}\mathbf{u}\right)=-\frac{\mbox{\ensuremath{\nabla}}P_{2}}{\rho} + F_{3d} + \nu\Delta\vv u,\label{eq:LANS-alpha-3D}\end{equation}
where $\mathbf{u}=\left(I-\alpha^{2}\Delta\right)\mathbf{u}_l$.\\

Applying the operator $\left(I-\alpha^{2}\Delta\right)$ to (\ref{eq:rational_model_3D}),
we obtain the equation verified by $\mathbf{u}=\left(I-\alpha^{2}\Delta\right)\mathbf{u}_l$
in the case of the Rational LES model:
\begin{equation}
\frac{\partial\mathbf{u}}{\partial t}-\mathbf{u}_l\boldsymbol{\times}\left(\nabla\boldsymbol{\times}\mathbf{u}\right)=-\frac{\mbox{\ensuremath{\nabla}}P_{3}}{\rho}+\mathbf{N}+ F_{3d} + \nu\Delta\vv u,\label{eq:rational_model_3D_v}\end{equation}
with $P_{3}=\left(I-\alpha^{2}\Delta\right)P_{2}$, and with
\[
\mathbf{N}=\alpha^{2}\left\{ -2\nabla\mathbf{u}_l^{T}.\nabla(\nabla\mathbf{u}_l)+\mathbf{u}_l\times\boldsymbol{\nabla}\times\left[\Delta\mathbf{u}_l\right]+\Delta\left[\mathbf{u}_l\times\left(\boldsymbol{\nabla}\times\mathbf{u}_l\right)\right]\right\} .\]
The two equations (\ref{eq:LANS-alpha-3D}) and (\ref{eq:rational_model_3D_v})
are equivalent if and only if $\mathbf{N}$ is a gradient, that is
if and only if $\nabla\times\mathbf{N}=0$. For two-dimensional vector-fields,
we have proven in the section above that $\delta M\mathbf{e}_{z}=\nabla\times\mathbf{N}=0$.
In contrast this is wrong in general for three-dimensional vector
fields, because of vortex-stretching type terms present for three
dimensional vector fields and non-present for two dimensional vector
fields. In order to prove this we give below an example of field
$\mathbf{u}_l$ for which $\nabla\times\mathbf{N}\neq0$.

Consider for example $\mathbf{u}_l=y^{2}\mathbf{e}_{x}-xz\mathbf{e}_{y}+xy\mathbf{e}_{z}$.
Then $\mathbf{u}_l$\footnote{As previously discussed, the spatial filter here
 is the rational approximation to the
  Gaussian filter as given in (3.12) or equivalently the Helmholtz
  filter (3.13). The unfiltered velocity is given by the inversion
  (deconvolution) of the above filter. Although one does
  not have to invoke the filter itself, when using the turbulence
  model in an {\em a posteriori} sense since the evolution equations
  are written explicitly in terms of just the large-scale
  velocity, it is important to
  conducting {\em a priori} tests.} is actually non-divergent: $\nabla\cdot\mathbf{u}_l=0$. By direct computation, we have $\boldsymbol{\nabla}\times\left[\Delta\mathbf{u}_l\right]=0$,
$\mathbf{u}_l\times\boldsymbol{\nabla}\times\left[\Delta\mathbf{u}_l\right]=0$,
$\nabla\times\left[\nabla\mathbf{u}_l^{T}.\nabla(\nabla\mathbf{u}_l)\right]=2\mathbf{e}_{x}$,
$\nabla\times\left\{ \Delta\left[\mathbf{u}_l\times\left(\boldsymbol{\nabla}\times\mathbf{u}_l\right)\right]\right\} =-8\mathbf{e}_{x}$
and then $\nabla\times\mathbf{N}=4\alpha^{2}\mathbf{e}_{x}\neq0$.

We thus conclude that the Rational LES and the LANS-$\alpha$ models are not equivalent in three dimensions.

\section{A numerical example}

The primary emphasis of this article is the above analytical
demonstration of the equivalence of the Rational LES model and the
LANS-$\alpha$ model in two dimensions rather than an evaluation of the performance of
the model(s) considered. Nevertheless, at the insistence
of one of the referees, we briefly present an example computation in
two dimensions in this section.

\begin{figure}
\includegraphics[width=6in]{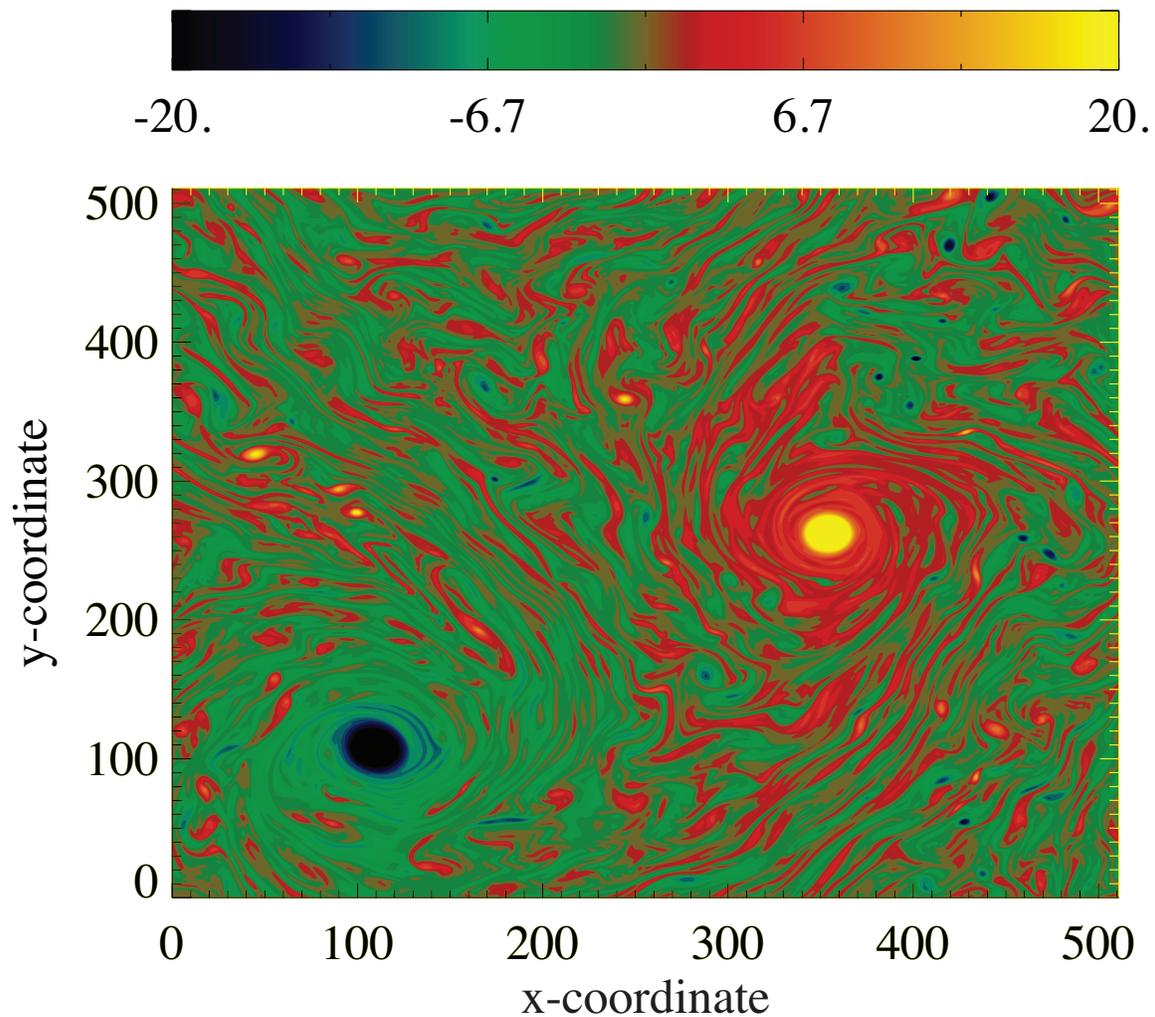}
\caption{Snaphot of vorticity field of reference run at statistical
  stationarity. The stochastic forcing is confined to a band of
  wavenumbers between 15 and 16 (domain is $2\pi \times 2\pi$) and
dissipation consists of a combination of Rayleigh friction and
hyperviscosity.}
\end{figure}

We consider a (stochastically) forced-dissipative flow in a doubly
period domain $2\pi$ on the side.  As is conventional in numerous
earlier investigations of two dimensional turbulence, dissipation $D$
consists of linear damping: $-r\omega$, where $r(=10^{-3})$ is a
frictional constant, and an eigth order hyperviscous term that acts as a
sink of the net-forward cascading enstrophy.  Forcing $F$ is scaled as
$F=\sqrt{2r}\tilde{F}$, where $\tilde{F}$ is an isotropic stochastic
forcing in a small band of wavenumbers $15\leq k_f< 16$ drawn from
an independent unit variance Gaussian distribution and which is
temporally uncorrelated: $\left<\tilde{F}_{\mathbf
    k}(t)\tilde{F}_{\mathbf k'}(t')\right>=\delta_{\mathbf k \mathbf
  k'}(t-t')$.  A fully-dealiased pseudo spectral spatial
discretization is used in conjunction with an adaptive fourth-fifth
order Runge-Kutta Cash-Karp time stepping scheme. The time step used
ensures that the relative error of the time increment is less that
$10^{-6}$, and with the time step ending up being much smaller than
that required by stability requirements.

\begin{figure}
\includegraphics[width=6in]{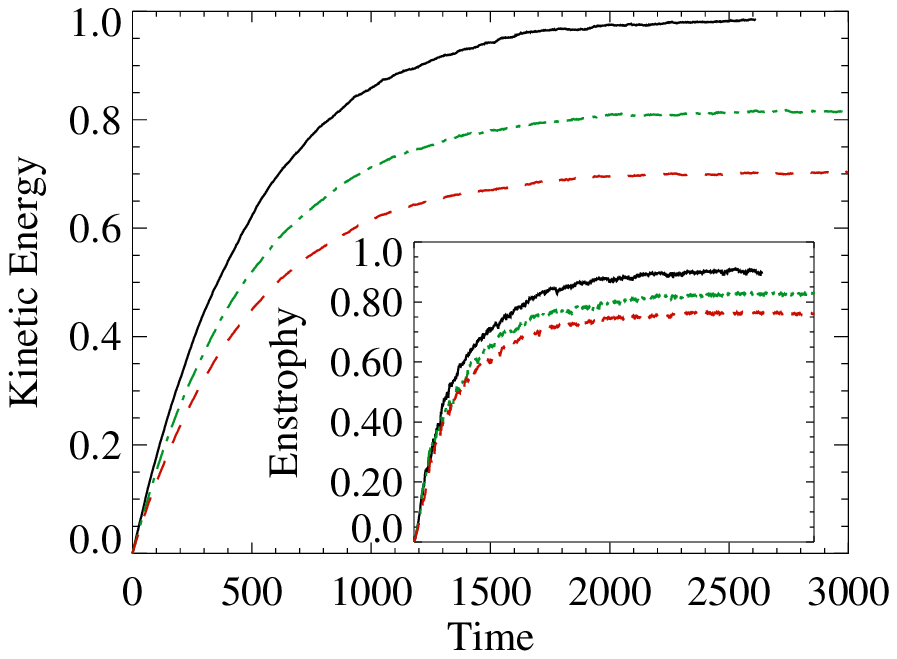}
\caption{The evolution of kinetic energy and enstrophy (inset) in the
  reference run (solid line; black), bare truncation run (dashed line;
  red) and the LES run (dot-dashed line; green). The horizontal axis
  in the inset spans the same range of times as in the main plot. The
  former uses a 512x512 physical space grid where as the latter two
  runs use a 128x128 grid. The changes introduced by the subgrid model are so as to
  improve the bare truncation run in the direction of the reference
  run.}
\end{figure}

For the reference computation, a 512x512 physical grid is chosen
giving a gridsize of $\pi/256$.  Figure 1 shows the vorticity field
after the flow has equilibrated (at $t$=2600 eddy turnover times.)
Given the stochastic forcing and the turbulent nature of the flow, a
statistical consideration of the flow is in order: The evolution of
the domain integrated kinetic energy and enstrophy as a function of
time is shown in Fig. 2. Figure 3 shows the one-dimensional spectral
density of kinetic energy and the spectral flux of kinetic energy
($\int_0^k 2\pi k \,dk \, \hbox{Re}(\wh{\vv u}^* \vv{\cdot}\wh{\vv{u}
  \vv{\cdot} \vv{\nabla} \vv{u}})$ where $\hbox{Re}(.)$ denotes the
real part, $\wh{.}$ denotes the Fourier transform and superscript $*$
denotes complex conjugate.) For the spectral flux of kinetic energy
diagnostic, we verify that the integral over all wavenumbers goes to zero
to within roundoff. Note that the x-axis in Fig. 3 is truncated at
wavenumber 60 to better focus on the range of scales of interest. In
all these figures, the solid line represents the reference run.

Next, we choose a filter width of $\pi/32$, and following arguments
similar to those in section 13.2 of Pope, 2000~\cite{pope-00}, we
choose an LES gridsize of $\pi/64$ (that corresponds to a 128x128
physical grid.) On the (coarser) LES grid, we perform two
simulations: One that we call a bare truncation---(\ref{eq:rational}),
but without the first term on the right hand side--and a second one
with the LES model discussed above---(\ref{eq:rational})---with the
rest of the setup being identical. The evolution of domain-integrated
kinetic energy and enstrophy and the spectral density and spectral
flux density of kinetic energy for these two simulations are shown
again in Figs. 2 and 3. The bare truncation run in these figures is
shown by dashed (red) lines, whereas the Rational LES or LANS-$\alpha$
model runs are shown by dot-dashed (green) lines.

\begin{figure}
\includegraphics[width=6in]{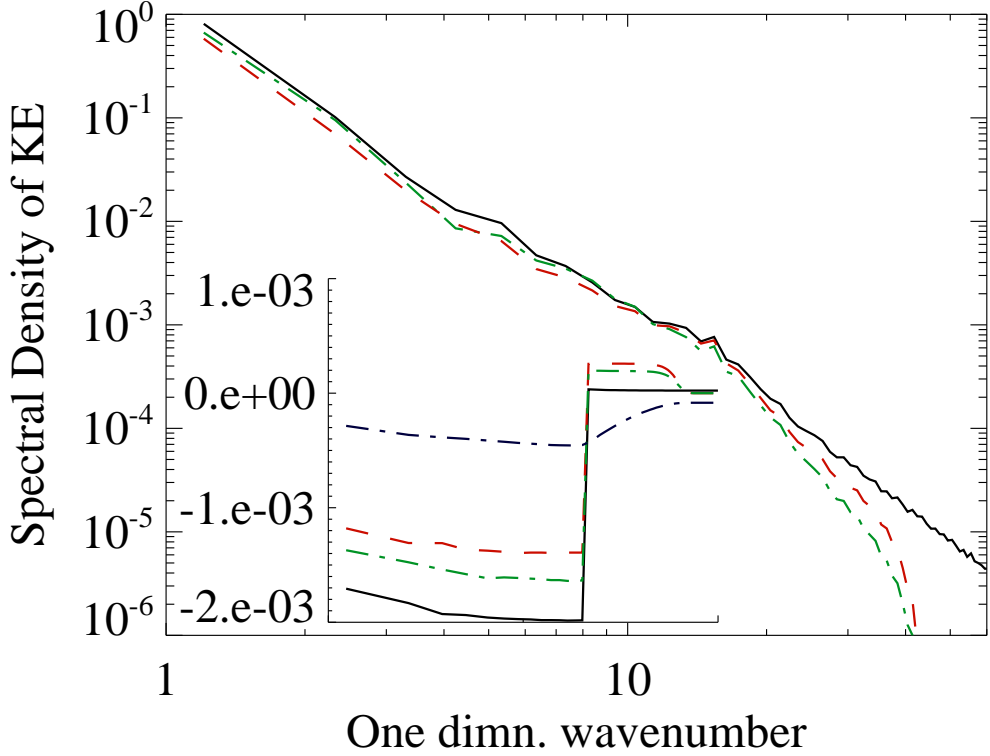}
\caption{The one-dimensional power spectral density (logarithmic scale) and spectral flux
  density (inset; linear scale) plotted as a function of the one-dimensional
  wavenumber (logarithmic scale). The horizontal axis in the inset spans the same range of
  wavenumbers as in the main plot. Reference run: solid line, black;
  bare truncation run: dashed line, red; and the LES run: dot-dashed
  line, green. The thick dot-dashed line in blue in the inset
  corresponds to the spectral flux of energy due to the subgrid
  model. In the range of scales where there is an inverse cascade of
  energy, the LES run is more energetic than the bare truncation run,
  and the LES run closely follows the reference run. However at the
  small scales, the energy spectrum of the LES run falls off faster
  than the bare truncation run (at these range of scales, the
  reference run is still inertial.) The increased level of energy at
  the large scales in the LES run is seen as due to a secondary
  inverse cascade that is put in place by the subgrid model
  (backscatter). In effect, as compared to the bare truncation run, in
  the LES run, the forward cascade of energy is reduced and the
  inverse cascade of energy is augmented.}
\end{figure}

In each of these diagnostics, the tendency of the model to improve on
the bare truncation is evident. In the spectral density plot, on
comparing the bare truncation with the model simulation, the tendency
of the model to deemphasize the small scales while increasing the
energy in the large scales is seen. The dynamics of how this is
achieved is seen to be that of an augmentation of the inverse cascade
by the model term as indicated by the blue line in the spectral flux
inset. The net result is that the full nonlinear flux of energy shows
a smaller forward cascade\footnote{The forward cascade is an artifact
  of finite resolution.  For details see \cite{nadiga-straub-10}} and
an increased inverse cascade, as compared to the bare truncation
simulation. And these changes in the spectral flux of energy are in
the direction of bridging the (coarser) bare truncation run to the
reference simulation. We note that a) we did not tune any of the
parameters to match the reference run; we anticipate that with tuning,
the LES results could better match the reference run, and that b) the
computations on the LES grid are about 60 to 100 times less
computationally intensive as compared to the reference run, with the
overhead for the model (over bare truncation) being negligible.

\section{Conclusion}
In its popular form, the LES approach to modeling turbulence comprises
of applying a filter to the original set of equations; the nonlinear
terms then give rise to unclosed residual terms that are then
modeled. However, the regularization approach to modeling turbulence
consists of, besides other possible considerations, a modification of
the nonlinear term based on filtering of one of the fields. The latter
approach, however, implies a model of the unclosed residual terms when
viewed from the point of view of the former. We consider the Rational
LES model~\cite{galdi-00} that falls under the former approach, and
the LANS-$\alpha$ model~\cite{holm-98} that falls under the latter
approach. In this article, we demonstrate that the two models are
equivalent in two dimensions, but not in three dimensions. Their
equivalence in two dimensions allows arguments about the mathematical
structure and physical phenomenology of either of the models to be
equally valid for the other.

\begin{acknowledgments}
This work was carried out under the LDRD-ER program (20110150ER) 
of the Los Alamos National Laboratory.
\end{acknowledgments}

\section{Appendix}

We derive in this appendix calculus identities for two-dimensional
or three-dimensional vector fields. We define for any vector fields
$\mathbf{A}$ and scalar $B$: $\nabla\mathbf{A}^{T}.\nabla(\nabla B)\equiv\partial_{i}A_{j}\partial_{ij}B$
(sum over repeated indices, here and in the following) and $\Delta\mathbf{A}^{T}.\nabla B=\Delta A_{j}\partial_{j}B$.
Similarly, for any two vectors fields $\mathbf{A}$ and $\mathbf{B}$,
we define $\nabla\mathbf{A}^{T}.\nabla(\nabla\mathbf{B})\equiv\partial_{i}A_{j}\partial_{ij}\mathbf{B}$
and $\Delta\mathbf{A}^{T}.\nabla\mathbf{B}=\Delta A_{j}\partial_{j}\mathbf{B}$
\begin{enumerate}
\item We first note that the useful, and easily derived, vector calculus identity\begin{equation}
\Delta\left(\mathbf{A}.\nabla B\right)=\Delta\mathbf{A}^{T}.\nabla B+2\nabla\mathbf{A}^{T}.\nabla(\nabla B)+\mathbf{A}.\nabla\left(\Delta B\right).\label{eq:Identity1}\end{equation}
\item We then note that for a 2D solenoidal vector field $\mathbf{u}$
($\nabla.\mathbf{u}=0$), if $\omega\equiv\left(\nabla\times\mathbf{u}\right).\mathbf{e}_{z}$,
then \begin{equation}
\Delta\mathbf{\mathbf{u}}^{T}.\nabla\omega=0.\label{eq:Identity2}\end{equation}
Indeed using the flow incompressibility $\nabla.\mathbf{u}=0$, we
have $\partial_{x}\omega=\Delta u_{y}$ and $\partial_{y}\omega=-\Delta u_{x}$,
then $\Delta\mathbf{\mathbf{u}}^{T}.\nabla\omega=\Delta u_{x}\partial_{x}\omega+\Delta u_{x}\partial_{x}\omega=0$.
\item Then for a 2D incompressible vector field $\mathbf{u}$ with $\omega=\left(\nabla\times\mathbf{u}\right).\mathbf{e}_{z}$,
using (\ref{eq:Identity1}) and (\ref{eq:Identity2}) we obtain \begin{equation}
\Delta\left(\mathbf{u}.\nabla\omega\right)=2\nabla\mathbf{u}^{T}.\nabla(\nabla\omega)+\mathbf{u}.\nabla\left(\Delta\omega\right).\label{eq:Identity3}\end{equation}

\remove{\item For any two vectors fields $\mathbf{A}$ and $\mathbf{B}$, identity
(\ref{eq:Identity1}) being true for $\mathbf{A}$ and any of the
components of $\mathbf{B}$, we have \begin{equation}
\Delta\left(\mathbf{A}.\nabla\mathbf{B}\right)=\Delta\mathbf{A}^{T}.\nabla\mathbf{B}+2\nabla\mathbf{A}^{T}.\nabla(\nabla\mathbf{B})+\mathbf{A}.\nabla\left(\Delta\mathbf{B}\right).\label{eq:Identity4}\end{equation}
}
\end{enumerate}


\listoffigures
\end{document}